\documentclass[twocolumn,aps,showpacs,preprintnumbers,amsmath,amssymb]{revtex4}
\usepackage{graphicx}
\usepackage{amsmath}
\usepackage{array}
\begin{document}
\title{Realization of quantum mechanical weak values of observables using entangled photons}
\author{G. S. Agarwal and P. K. Pathak} \affiliation{Department of Physics, Oklahoma
State University, Stillwater, Oklahoma 74078, USA}
\date{\today}
\begin{abstract}
We present a scheme for realization of quantum mechanical weak
values of observables using entangled photons produced in parametric
down conversion. We consider the case when the signal and idler
modes are respectively in a coherent state and vacuum. We use a low
efficiency detector to detect the photons in the idler mode.This
weak detection leads to a large displacement and fluctuations in the
signal field's quantum state which can be studied by monitoring the
photon number and quadrature distributions.
\end{abstract}
\pacs{03.65.Ta, 03.67.-a, 42.50.Xa} \maketitle In a remarkable paper
Aharonov, Albert and Vaidman \cite{aharonov} discovered that under
conditions of very weak coupling between the system and the
measuring device the uncertainty in a single measurement could be
very large compared with the separation between the eigenvalues of
the observable. This is quite a departure from the usual idea of
projective measurements where the measurement projects on to one of
the eigenvalues \cite{VonNeumann}. Aharonov et al proposal was
further clarified by several
authors\cite{duck,johansen,knight,ruseckas}. In particular Duck {\it
et al.} \cite{duck} also suggested an optical experiment to verify
the idea. Such an optical experiment was performed by Ritchie {\it
et al.} \cite{ritchie}. This experiment uses the classical light
source like a laser beam; a birefringent medium and the pre and post
selection of the polarization of the transmitted beam. The
experiment is well explained using classical physics as discussed in
the paper by Ritchie {\it et al} \cite{ritchie}. It would therefore
be interesting to find situations which are strictly quantum in
nature even though the experiment of Ritchie {\it et al.} was
repeated at single photon level and in particular weak values of
photon arrival times were measured \cite{ahnert}. Further weak
values of the polarization of a single photon were reported
\cite{wiaeman2}. In this letter we show how the idea of Aharonov et
al could be realized using entangled photons produced in the process
of parametric down conversion thus making their proposal within the
reach of current experiments. We specifically discuss the weak
values associated with the measurement of the photon number and the
quadrature of the signal field. We also discuss how the weak values
reflect in the fluctuations of the state of the signal field. Our
explicit result for the state of the signal field shows the role of
quantum interferences in the weak values of the observables. We
require two ingredients for the observation-high value of the
squeezing parameter and well controlled detector. Both of these are
feasible. We note that Zambra et al \cite{zambra} have very
successfully demonstrated the application of avalanche detectors for
the reconstruction of photon statistics. Further in recent
experiments \cite{caminati} high values of the squeezing parameter
have been achieved. We note [cf Eqs.(\ref{peak}), (\ref{width}),
(\ref{meann})] that if the squeezing is not very high then we need
to use smaller detection efficiencies. Our proposal contrasts the
previous ones \cite{ritchie,gisin} which used classical Gaussian
beams and birefringent medium.

\begin{figure}
\centering
\includegraphics[width=3in]{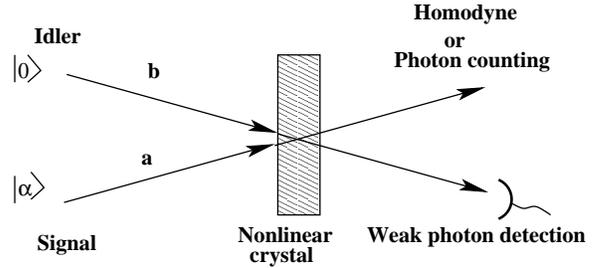}
\caption{The schematic arrangement for realization of weak values
 using entangled photons. The idler photon is detected with varying efficiency.}
\label{fig1}
\end{figure}

In Fig. \ref{fig1}, we show a schematic arrangement for
realization of the weak values using entangled photons. The
entangled photons are generated by an optical parametric amplifier
(OPA). In OPA the pump field interacts nonlinearly in an optical
crystal having second order nonlinearity. As a result of
annihilation of one photon of pump field two entangled photons
propagating in two different directions are generated
simultaneously. In our scheme, we consider that the signal mode is
initially in a coherent state $|\alpha\rangle$ while the idler is
initially in vacuum. The photons generated in signal mode produce
excitation \cite{tara} in coherent field $|\alpha\rangle$
presented initially. If $n$-photons are generated in the idler
mode, the state of the signal mode will be in $n$-photon added
coherent state. In our scheme, we perform weak detection of the
idler photons by using a low efficiency detector which requires a
large number of photons to be produced in the idler mode. Thus we
consider OPA working under high gain conditions.
We note that a similar arrangement with OPA under low gain
conditions has been used in a recent experiment by Zavatta et al
\cite{Zavatta}, for generating photon added coherent states
\cite{tara}. Using the interaction Hamiltonian for the OPA and
under the assumption of no pump depletion, the state of the
outgoing signal and idler fields can be written as
\begin{equation}
|\psi\rangle=\exp\left(ra^{\dag}b^{\dag}-rab\right)|\alpha,0\rangle\label{evol}
\end{equation}
where  $r$ is gain of the amplifier. Using the Baker Campbell
Hausdorff identity, the Eqn (\ref{evol}) simplifies to
\begin{eqnarray}
|\psi\rangle&=&\exp\left(\tanh
ra^{\dag}b^{\dag}\right)\exp\left[-\ln\cosh r
(a^{\dag}a+b^{\dag}b+1)\right]\nonumber\\&&\times\exp\left(-\tanh
rab\right)|\alpha,0\rangle\nonumber\\
&=&\frac{e^{-\frac{1}{2}|\alpha|^2\tanh^2r}}{\cosh r}\exp\left(\tanh
ra^{\dag}b^{\dag}\right)|\alpha',0\rangle \label{eq3}
\end{eqnarray}
where $\alpha'=\alpha/\cosh r$. The Eq (\ref{eq3}) shows how the OPA
generates correlated pair of photons one in signal mode and one in
idler mode simultaneously. According to the von Neumann postulate
the measurement of the state of the idler mode in $n$-photon Fock
state $|n\rangle$ would project the state of the signal mode in
$n$-photon added coherent state $a^{\dag n}|\alpha'\rangle$. However
we now follow the idea of Aharonov et al on weak measurements. We
measure idler field weakly, i.e. the measurement does not make the
idler field to collapse in a single Fock state, with definite number
of photons, but a probabilistic mixture of various Fock states of
different number of photons. The weak detection is performed by a
low efficiency detector \cite{footnote}. Clearly, the weak
measurement of the idler field will project the signal field in a
superposition of various photon added coherent states\cite{tara}. We
would now show how the weak detection of idler photons leads to
unexpectedly large values of signal field. The density matrix for
signal-idler fields is
\begin{equation}
\rho=|\psi\rangle\langle\psi|, \label{density}
\end{equation}
where $|\psi\rangle$ is given by Eq.(\ref{eq3}). We detect idler
field in the $n$-photon Fock state $|n\rangle$ by using a detector
of quantum efficiency $\eta$. The projected state of the signal
field is
\begin{equation}
\rho_{n}^{(s)}=A\sum_{m=n}^{\infty}\left(\begin{array}{c}m\\n\end{array}\right)
\eta^n(1-\eta)^{n-m}\langle m|\rho|m\rangle, \label{detect}
\end{equation}
where $A$ is normalization constant. Using (\ref{density}) and
(\ref{eq3}), Eq (\ref{detect}) takes the form
\begin{equation}
\rho_{n}^{(s)}=N\sum_{m=n}^{\infty}\left(\begin{array}{c}m\\n\end{array}\right)
\eta^n(1-\eta)^{n-m}\frac{\tanh^{2m}r}{m!}a^{\dag
m}|\alpha'\rangle\langle\alpha'|a^{m}, \label{signal}
\end{equation}
where $N$ is new normalization constant and the constant term
$e^{-|\alpha|^2\tanh^2r}/\cosh^2 r$ has been absorbed in $N$.
\begin{figure}
\includegraphics[width=3in]{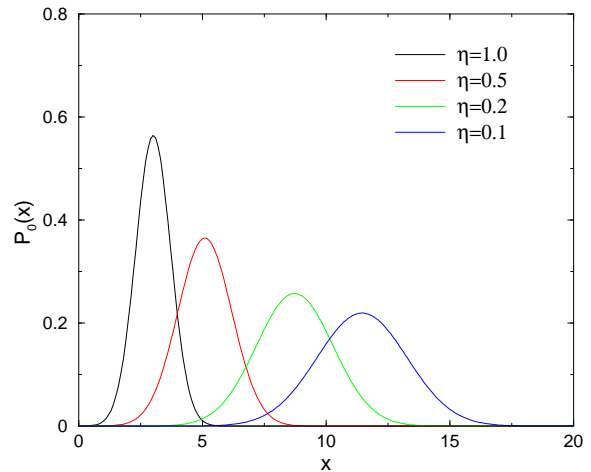}
\caption{\label{fig2}(Color online) The quadrature distribution of
signal field after measuring idler photons in vacuum state by using
detector of efficiency $\eta$.}
\end{figure}
From Eq (\ref{signal}), it is clear that because of non-unity
quantum efficiency of the detector, the measurement of the idler
field can not project the signal field in one of its eigenstate. The
projected state of the signal field is a superposition of various
eigenstates. Further, for smaller values of $\eta$ and larger values
of OPA gain parameter $r$, many eigenstates in the superposition
contributes significantly.

It should be noted here, that in our scheme we do not perform
measurement on a two state system as discussed by Aharonov et al
\cite{aharonov} in their original proposal. We perform measurement
on an infinite dimensional system. Here we are particularly
interested in detecting the idler field in vacuum state. Note that
the detection of idler in vacuum state for a range of values of the
efficiency is enough to reconstruct the full idler field
\cite{zambra}. From Eq.(\ref{signal}), the weak measurement of the
idler field in the state $|0\rangle$, projects the signal field in
the state
\begin{equation}
\rho_{0}^{(s)}=N\sum_{m=0}^{\infty}(1-\eta)^{m}\frac{\tanh^{2m}r}{m!}a^{\dag
m}|\alpha'\rangle\langle\alpha'|a^{m}. \label{vacsig}
\end{equation}
The above conditional state of the signal field can be measured
either through the photon number distribution or via the
quadrature distribution.We next calculate these and show how the
weak values get reflected in such distributions.

The quadrature distribution of the projected signal field
(\ref{vacsig}), when idler field is detected in the vacuum state by
using detector of quantum efficiency $\eta$, is given by
\begin{equation}
P_{0}(x)=N\sum_{m=0}^{\infty}(1-\eta)^{m}\frac{\tanh^{2m}r}{m!}\langle
x|a^{\dag m}|\alpha'\rangle\langle\alpha'|a^{m}|x\rangle,
\label{xvac}
\end{equation}
where $|x\rangle$ is eigenstate in the quadrature space.A long
calculation leads to the following compact expression for the
quadrature distribution of the signal field
\begin{equation}
P_0(x)=\sqrt{\frac{1-\epsilon}{\pi(1+\epsilon)}}\exp\left[-\frac{1-\epsilon}{1+\epsilon}
\left(x-\frac{\sqrt{2}\alpha'}{1-\epsilon}\right)\right]
\label{quad}
\end{equation}
where $\epsilon=(1-\eta)\tanh^2r$. From Eq (\ref{quad}), it is clear
that the projected state of the signal field (\ref{vacsig}) has
Gaussian quadrature distribution. The peak of the distribution
appears at
\begin{equation}
x=\sqrt{2}\alpha'/(1-\epsilon), \label{peak}
\end{equation}
and the width of the distribution $\delta x$ is given by
\begin{equation}
2(\delta x)^2= \frac{1+\epsilon}{1-\epsilon}.\label{width}
\end{equation}
It is clear from Eq (\ref{peak}) and Eq (\ref{width}) that for low
efficiency detector $(\eta<<1)$ and high gain OPA $(r>1)$, as the
value of $\epsilon$ tends towards $1$, the peak in the quadrature
distribution occurs for exceptionally large values of $x$.
Further, the width of the distribution also becomes very large for
these values of the parameters.Interestingly enough in our model
the width of the distribution also depends on the weakness of the
measurement.

In Fig.\ref{fig2}, we show the quadrature distributions of the
projected states of the signal field after detecting the idler
field in vacuum state. For 100\% detection efficiency the maxima
in the quadrature distributions occurs at $x\approx3$
corresponding to the value $x=\sqrt{2}\alpha'$, where
$\alpha'=\alpha/\cosh r$ and $\alpha=5$ and $r=1.5$. We find that
for lower detection efficiency the maxima in the quadrature
distribution shifts to the exceptionally larger values of $x$. For
$\eta=0.1$ the maxima in x-quadrature appears around $x\approx12$.
Further, the spread in the distribution becomes very large for
such smaller values of $\eta$. This is a remarkable realization of
the idea of Aharonov et al using entangled photons.

\begin{figure}
\includegraphics[width=3in]{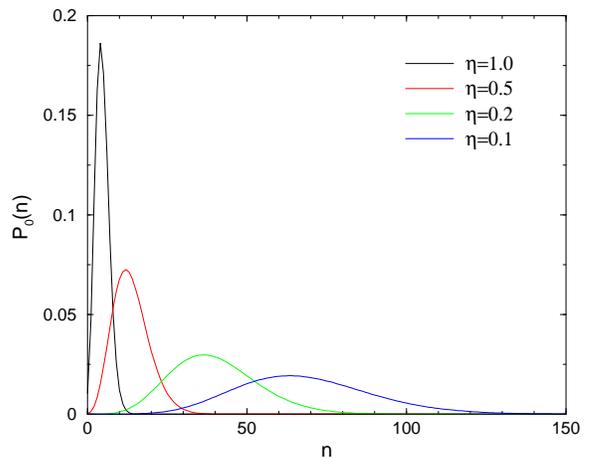}
\caption{\label{fig3}(Color online) The photon distribution of
signal field after measuring idler photons by using detector of
efficiency $\eta$. The   idler field is detected in vacuum state
$|0\rangle$}
\end{figure}

In order to understand the exact nature of the weak values we look
at Eq(\ref{vacsig}) for the projected state of the signal field.
Clearly, the projected state of the signal field is superposition
of photon added coherent states $a^{\dag m}|\alpha'\rangle$
generated by successive addition of the photons in the signal
mode. The amplitude of the $m$-th term in Eq(\ref{vacsig}) is
proportional to $\epsilon^m/m!$, where
$\epsilon=(1-\eta)\tanh^2r$. For $\eta=0.1$ and $r=1.5$, as the
value of $\epsilon$ is $0.74$, the amplitude of the fifth term is
of the order of $10^{-3}$. Further the higher order terms will
have much smaller amplitude and can be neglected. The quadrature
distribution of $5$-photon added coherent state
$a^{\dag5}|\alpha'\rangle$ will have maxima at
$x\approx\sqrt{2}\sqrt{|\alpha'|^2+5}$. Thus the highest order
contributing eigen state of the signal field has maxima at
$x\approx4.5$. In Fig.\ref{fig3} the maxima in the quadrature
distribution corresponding to these parameters occurs at
$x\approx12$, which is exceptionally large and there is no doubt
that the projected values of the signal field in our scheme are
weak values. The exceptional displacement in the maxima of the
quadrature distribution occurs as a result of interferences
between various states contributing to the projected state of the
signal field. Further it should also be noted that the photon
added coherent states $a^{\dag m}|\alpha'\rangle$ show more and
more squeezing in their quadrature on increasing $m$ \cite{tara},
while the projected state of the signal field exhibits broadening
in the quadrature distribution. Clearly choosing smaller detection
efficiencies (few \%) which are definitely feasible \cite{zambra}
would lead to larger displacement and large fluctuations.

Next we show how the weak measurements get reflected in the photon
number distribution of the signal field.The photon distribution of
the projected state of the signal field (\ref{vacsig}) is
calculated as follows
\begin{equation}
P_{0}(n)=N\sum_{m=0}^{\infty}(1-\eta)^{m}\frac{\tanh^{2m}r}{m!}\langle
n|a^{\dag m}|\alpha'\rangle\langle\alpha'|a^{m}|n\rangle,
\label{nvac}
\end{equation}
\begin{eqnarray}
P_{0}(n)&=&N(1-\eta)^n\tanh^{2n}r\times\nonumber\\
&\sum_{m=0}^{n}&\frac{n!}{m!(n-m)!^2}
\left(\frac{\alpha'^{2}}{(1-\eta)\tanh^2r}\right)^{n-m}.
\label{new2}
\end{eqnarray}
Using definition of Laguerre polynomials and evaluating the
normalization constant, the Eq (\ref{new2}) takes the form
\begin{equation}
P_0(n)=(1-\epsilon)e^{-\alpha'^2/(1-\epsilon)}
\epsilon^nL_n\left(-\frac{\alpha'^2}{\epsilon}\right)
\label{normnvac}
\end{equation}
where $L_n(q)$ is Laguerre polynomial of order $n$.
The photon distributions of the projected signal state
(\ref{vacsig}) are shown in Fig.\ref{fig3}. For unity detection
efficiency the field has maxima at $n\approx5$ corresponding to the
coherent state $|\alpha'\rangle$. As the detection efficiency $\eta$
decreases the peak in the distribution moves very fast towards the
higher values of $n$ and the width of the distribution also
increases. As we have discussed earlier, for $\eta=0.1$, only terms
up to $m=5$ can contribute significantly in the projected signal
state (\ref{vacsig}). Thus the signal field contains its highest
order eigenstate having maxima in the photon distribution at
$|\alpha'|^2+5\approx10$. But the actual weak value of the maximum
photon numbers in the distribution occurs at $n\approx70$.

For further probing the field statistics of the projected signal
states in weak measurement, we calculate the Mandel Q-parameter
defined by
\begin{equation}
Q=\frac{\langle n^2\rangle-\langle n\rangle^2}{\langle n\rangle}-1
\label{defmandel}
\end{equation}
where $\langle n\rangle$ is average number of photons in the
projected state of the signal field. The average number of photons
for state (\ref{vacsig}) is
\begin{equation}
\langle
n\rangle=\frac{\epsilon-\epsilon^2+\alpha'^2}{(1-\epsilon)^2}
\label{meann}
\end{equation}
It is clear from Eq (\ref{meann}) that the average number of
photons becomes very large for $\epsilon\rightarrow1$ for smaller
value of $\eta$. The calculated value of Mandel Q-parameter for
the state (\ref{vacsig}) is
\begin{equation}
Q=\frac{(\epsilon-\epsilon^2+\alpha'^2)(1-\epsilon+\alpha'^2)-\alpha'^4}
{(1-\epsilon)^2(\epsilon-\epsilon^2+\alpha'^2)}-1 \label{mandel}
\end{equation}
\begin{figure}
\includegraphics[width=3in]{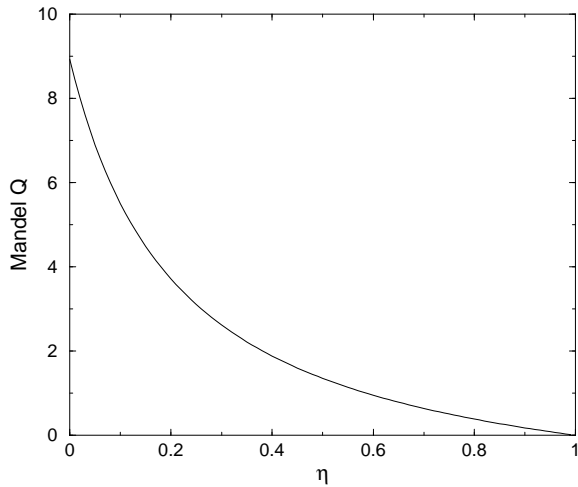}
\caption{The value of
Mandel's Q-parameter for the signal field after detecting idler
field in vacuum state}. \label{fig4}
\end{figure}
In Fig.\ref{fig4}, we plot Mandel Q-parameter for the signal
state(\ref{vacsig}) with respect to the efficiency of the detector
used to measure the idler field. For smaller values of detection
efficiency $Q$-parameter has large positive values and the states
of the signal field follow super-Poissonian statistics. As the
detector efficiency increases the value of $Q$-parameter
decreases. For the detector efficiency more than $0.9$
$Q$-parameter for the state (\ref{vacsig}) is zero which reflects
that the projected state of the signal field is coherent state
$|\alpha'\rangle$.

Next we calculate the Wigner distribution of the projected states
of the signal field. The Wigner distribution for the state having
density matrix $\rho$ can be obtained using coherent states from
the formula
\begin{eqnarray}
W(\gamma)=\frac{2}{\pi^{2}}e^{2|\gamma|^{2}}\int \langle
-\beta|\rho|\beta\rangle
e^{-2(\beta\gamma^{*}-\beta^{*}\gamma)}d^{2}\beta. \label{defn}
\end{eqnarray}
For state (\ref{vacsig}) the Wigner function is found to be
\begin{equation}
W_0(\gamma)=\frac{2(1-\epsilon)}{\pi(1+\epsilon)}
\exp\left[-\frac{2(1-\epsilon)}{1+\epsilon}\left|\gamma-\frac{\alpha'}{1-\epsilon}\right|^2\right]
\label{vacwign}
\end{equation}
The Wigner distribution of the state (\ref{vacsig}) is Gaussian
whose width is greater than the width of the distribution
associated with a coherent state. Hence the Glauber-Sudarshan
distribution is also well defined Gaussian with a width
$\epsilon/(1-\epsilon)$ and centered at
$\alpha\prime/(1-\epsilon)$. The Wigner function shifts to larger
values and broadens as the detection efficiency goes down.

In conclusion we have shown how one can use entangled photon pairs
produced in a high gain parametric amplifier and imperfect
measurements on the idler field to realize the idea of weak values
of the observable at the level of quantized fields. We show how the
weak measurements of the idler field produce exceptionally large
changes in the quantum state of the signal field. We show large
changes in both mean values of the observable as well as in the
fluctuations. For illustration purpose we have chosen to detect the
idler field in vacuum state.We could choose to measure the idler in
some other state. This would lead to similar results. We add that
detection of the idler in single photon state produces nonclassical
character of the signal field.

The authors thank NSF grant no. CCF 0524673 for supporting this
work. GSA also thanks Marco Bellini for interesting
correspondence.

\end{document}